\RequirePackage{fix-cm}
\documentclass[smallextended]{svjour3}       
\smartqed  
\usepackage{graphicx}
\begin{document}

\title{A Fast Response Mission to Rendezvous with an Interstellar Object
}


\author{Darren Garber         \and
        Louis D. Friedman  \and
        Artur R. Davoyan \and
        Slava G. Turyshev \and
        Nahum Melamed \and
        John McVey \and
        Todd F. Sheerin
}


\institute{Darren Garber  \at
                 NXTRAC, Rolling Hills Estates, CA 90274 USA
           \and
           Louis D. Friedman \at
           The Planetary Society, Pasadena, CA 91101 USA
            \and
           Artur Davoyan \at
           Mechanical and Aerospace Engineering Department, University of California, Los Angeles, CA 90095 USA
           \and    
           Slava G. Turyshev \at  
           Jet Propulsion Laboratory, California Institute of Technology,  Pasadena, CA 91109 USA
           \and   
           Nahum Melamed, John McVey, and Todd F. Sheerin
           \at
           The Aerospace Corporation, El Segundo, CA 90245 USA
}

\date{Received: date / Accepted: date}

\maketitle

\begin{abstract}
A solar sail propelled small satellite mission concept to intercept and potentially rendezvous with newly discovered transient interstellar objects (ISOs) is described.  The mission concept derives from the proposal for a technology demonstration mission for exiting the solar system at high velocity, eventually to reach the focal region of the solar gravitational lens.  The ISO mission concept is to fly a solar sail toward a holding orbit around the Sun and when the ISO orbit is confirmed, target the sailcraft to reach an escape velocity of over 6 AU/year.  This would permit rapid response to a new ISO discovery and an intercept within 10 AU from the Sun.  Two new proven interplanetary technologies are utilized to enable such a mission: i) interplanetary smallsats, such as those demonstrated by the MarCO mission, and ii) solar sails, such as demonstrated by LightSail and IKAROS missions and developed for NEA Scout and Solar Cruiser missions.  Current technology work suggests that already within this decade such a mission could fly and reach an ISO moving through the solar system.  It might enable the first encounter with an ISO to allow for imaging and spectroscopy, measurements of size and mass, potentially giving a unique information about the object's origin and composition. A similar approach may be used to allow for a sample return.

\end{abstract}

\section{Introduction}
\label{sec:ntro}

Two interstellar objects (ISOs) have been discovered passing through our inner solar system in the past five years: `Oumuamua discovered on October 19, 2017 \cite{Jewitt:2017,Mamajek:2017,Bannister:2019} and Borisov discovred on  August 30, 2019 \cite{Guzik:2020}.  As the first ever observations of an ISO, they are of significant scientific and public interest.  These two discoveries presage more, with some models predicting the transit of at least one such body per year through the inner regions of our solar system.  While we do not know the frequency of such interstellar visits, it is believed that there may be up $\sim10^4$ of similar ISOs inside the orbit of Neptune \cite{Jewitt:2017,Fuente-Marco:2018} and we can detect them at least once every 5 years with current capabilities (i.e., PanSTARRS\footnote{The Panoramic Survey Telescope and Rapid Response System (Pan-STARRS): https://www.ifa.hawaii.edu/research/Pan-STARRS.shtml}) and perhaps as frequently as once a year with the upcoming Vera C. Rubin Observatory\footnote{The Vera C. Rubin Observatory, formerly the Large-aperture Synoptic Survey Telescope (LSST), https://www.lsst.org/} in about 5 years.  To aid the search, new and more capable instruments and techniques are being developed. 

A space mission to encounter an ISO requires knowing the ISO's precise orbit\footnote{The uncertainty in the ISO's orbit is negligible for the design of the mission, but for the actual flight mission will be resolved and refined prior to launch and continuously during the rendezvous. Our approach is to be able to correct for significant offsets (or over 100s of arcsec in RA and DEC) of the target during the post perihelion passage outward flight.}, something likely to occur only months before it reaches its closest approach to the Sun and begins its exit from the solar system \cite{Siraj-Loeb:2019}.  Given previous examples of ISO speeds -- `Oumuamua departed our solar system at $\sim5$ AU/year, while Borisov exited with a speed of $\sim6.5$ AU/year -- it is unlikely that a spacecraft relying on conventional chemical propulsion capabilities will be capable of catching with or even matching an ISO's exit speed over relevant mission timescales and for overall program cost. For reference, the fastest spacecraft we have ever sent out of the solar system, Voyager, currently has a speed of $\sim3.57$ AU/year. 

Here, we propose a practical and potentially low-cost way to intercept or even rendezvous with an ISO with a total mission duration of $<5$ years. An even more ambitious option involves an ISO sample return with a total mission duration of $<$10 years. The concept derives from a Technology Demonstration Mission (TDM) identified in a study of a solar gravitational lens (SGL) and the mission to the SGL's focal region (SGLF) \cite{Turyshev-etal:2020-PhaseI,Turyshev-etal:2020-PhaseII}. The enabling technologies of this new approach are interplanetary smallsats and solar sails\footnote{The TDM is  designed to prove the basic concept that relies on interplanetary smallsats and solar sail control for the mission to the SGL's focal region.  Realizing that the TDM could achieve hyperbolic velocities $>6$ AU/year, the NIAC Phase III SGL team extended the concept to conceive the ISO science mission concept described here.}.  Both have now been proven in flight: Mars Cube One (MarCO), were the first interplanetary smallsats. Two solar sails have also flown successfully: LightSail \cite{Spencer:2021} in Earth Orbit and IKAROS \cite{Tsuda:2013} in interplanetary flight from Earth to Venus. Our approach is to use smallsat probes propelled by solar sails to fast exit velocities ($>5$ AU/year). Below, we discuss possible mission parameters and spacecraft architecture to realize this ISO mission capability.







\section{The Trajectory}
\label{sec:traject}

\begin{figure}
	\centering
	        \includegraphics[scale=0.70]{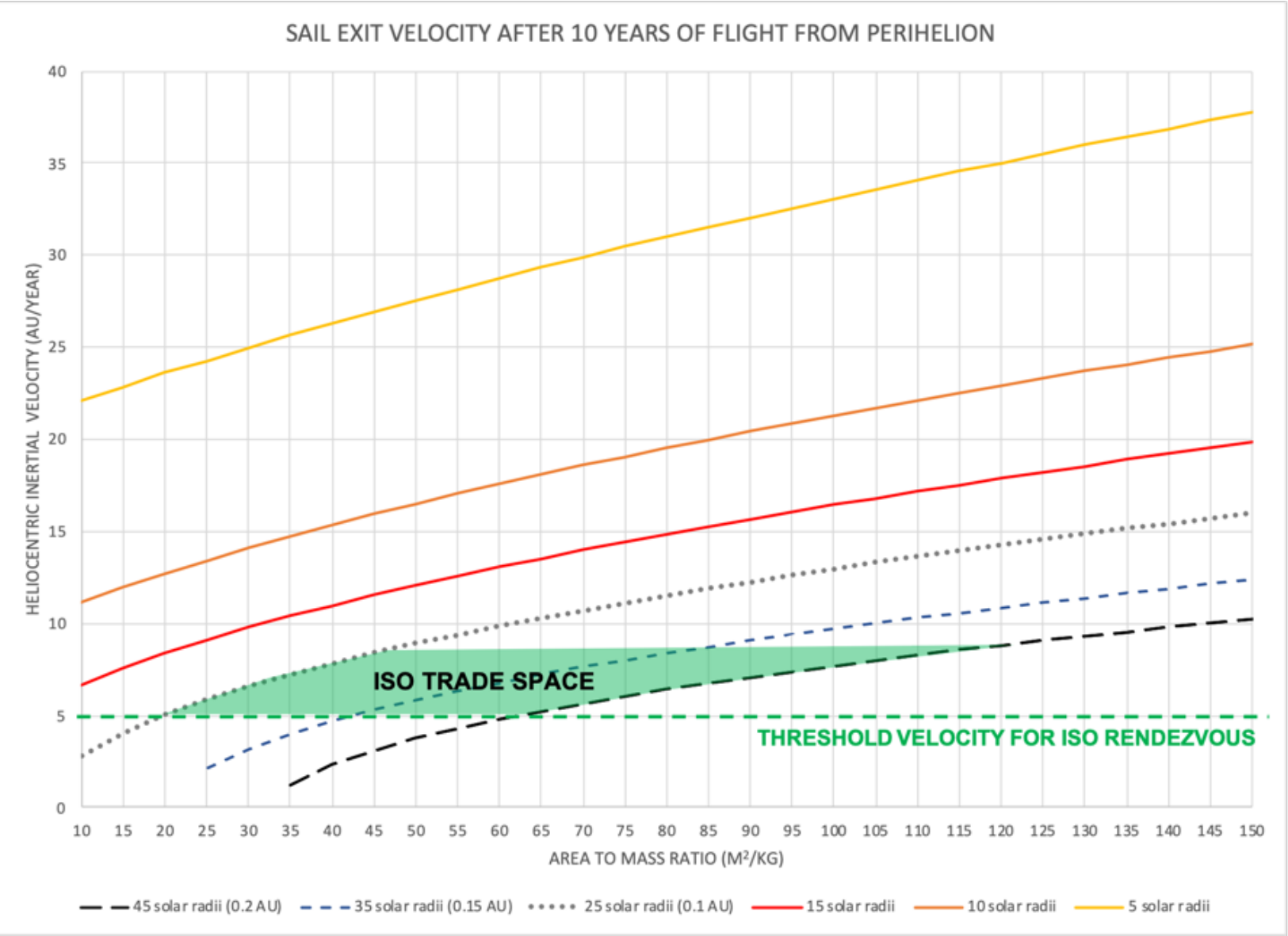}
	  \caption{Hyperbolic excess velocity vs. sailcraft's sail area-to-mass-ratio, $A/m$, for different perihelion distances.}
	  \label{fig1}
\vskip  5pt
	\centering
		\includegraphics[scale=0.70]{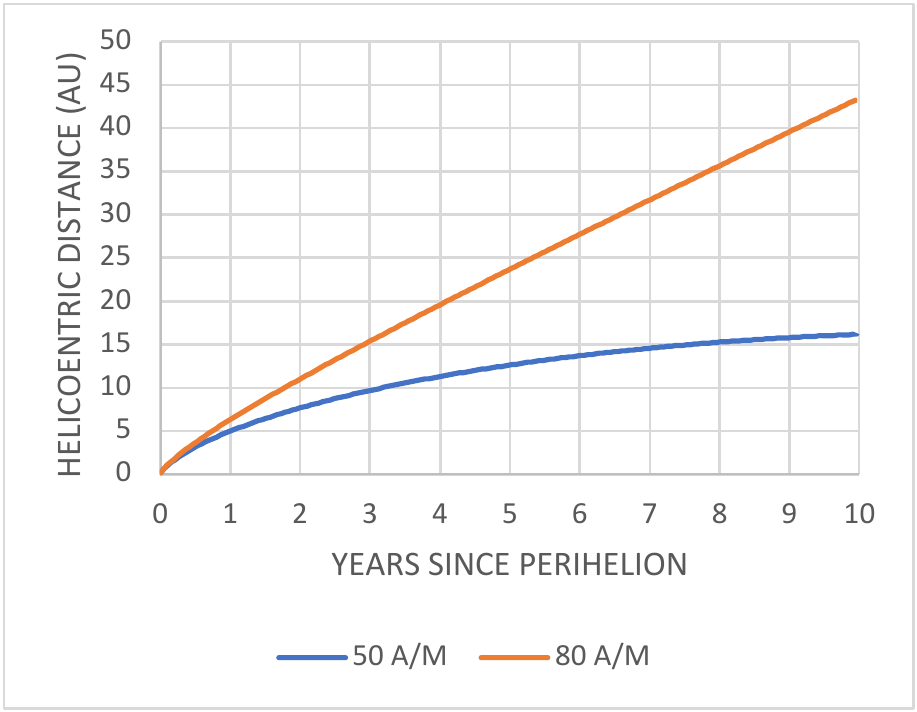}
	  \caption{Heliocentric distance vs. time from 0.2 AU perihelion for two sailcraft area-to-mass ratios, $A/m$.}
	  \label{fig2}
\end{figure}

Unlike conventional electric and chemical propulsion systems, solar sails use solar radiation pressure for thrust and are therefore not constrained by the limitations of the rocket equation. As several previous studies suggest (see \cite{Friedman-Garber:2014} and references therein), a sailcraft can be propelled around with sufficiently low perihelion and placed on hyperbolic trajectories with very high excess velocities ($v_{\rm inf}$) from the solar system. The hyperbolic velocity after solar perihelion depends on the perihelion distance, the sail area ($A$) and the total mass ($m$) of the sailcraft.  Figure~\ref{fig1} shows the corresponding tradeoff (see \cite{Friedman-Garber:2014,Friedman-Turyshev:2018} and references therein). Specifically, by flying to a perihelion of 0.2 AU away from the Sun, a sailcraft with an area-to-mass ratio of $A/m=60 \,{\rm m}^2/{\rm kg}$ can be accelerated to $v_{\rm inf} >5 ~{\rm AU/yr}$, which is comparable to that of an ISO. For closer perihelion, or for higher $A/m$ sailcraft, even higher velocities are possible. Notably, the bound on the maximum cruise velocity that a
solar sail may reach is given as \cite{Davoyan:21}
$$
v_{\rm inf}\simeq \sqrt{-\frac{2 \mu_\odot}{1{\rm AU}+d_0} +2(2r+\alpha)\frac{S_{\rm 1AU}}{d_0}\frac{A}{m}},
$$
where $\mu_\odot = G M_\odot$ is the solar
gravitational parameter with $G$ denoting gravitational constant
and $M_\odot$ solar mass, respectively, and $S_{\rm 1AU}$ is the solar irradiance at 1 AU, also $r$ and $\alpha$ are the optical properties of the sail (i.e., solar reflectivity and absorptivity), $A/m$ is the sail area to total spacecraft mass ratio, and $d_0$ being the solar perihelion distance.

Figure~\ref{fig2}  shows heliocentric distance from perihelion vs. time for a 0.2 AU perihelion pass and for two different values of $A/m$.  These profiles suggest that ISO interception and rendezvous missions may be implemented within 10~AU, significantly simplifying communication and tracking challenges. 
This figure  demonstrates that a spacecraft with a sail $A/m =80 \, {\rm m}^2/{\rm kg}$ reaches 5 AU in $\sim 9$ months, and 10 AU in $\sim 22$ months. The velocities at those distances are 6.7 AU/yr and 5.6 AU/yr, respectively. The hyperbolic excess velocity of this trajectory is $>4$~AU/yr, faster than Voyager.   Even with a heavier spacecraft assumption, for instance with $A/m = 50\, {\rm m}^2/{\rm kg}$, the times to 5 AU and 10 AU are 1 year and 3 years, respectively.  Even in this scenario, we will have caught the ISO.

Another important consideration for such a mission is timing. As the time window from ISO detection to the beginning of its outbound leg from the solar system is about several months based on the examples provided by `Oumuamua and Borisov, it is critically important to ensure that a spacecraft is built and ready to be launched with a very short lead time. The solar sailing approach introduced here is not sensitive to timing and affords rapid response. Our mission can start from $C_3=0$, i.e., at escape velocity from Earth, and may even be a rideshare payload on a lunar or other interplanetary mission.  From there, the sailcraft spirals in toward the Sun continuously optimizing its trajectory to reach a desired perihelion (distance and time) to place it on the intercept trajectory leaving the solar system. In this way, the mission could be launched in anticipation of and prior to an ISO discovery. 
 
\begin{figure}
	\centering
		\includegraphics[scale=0.70]{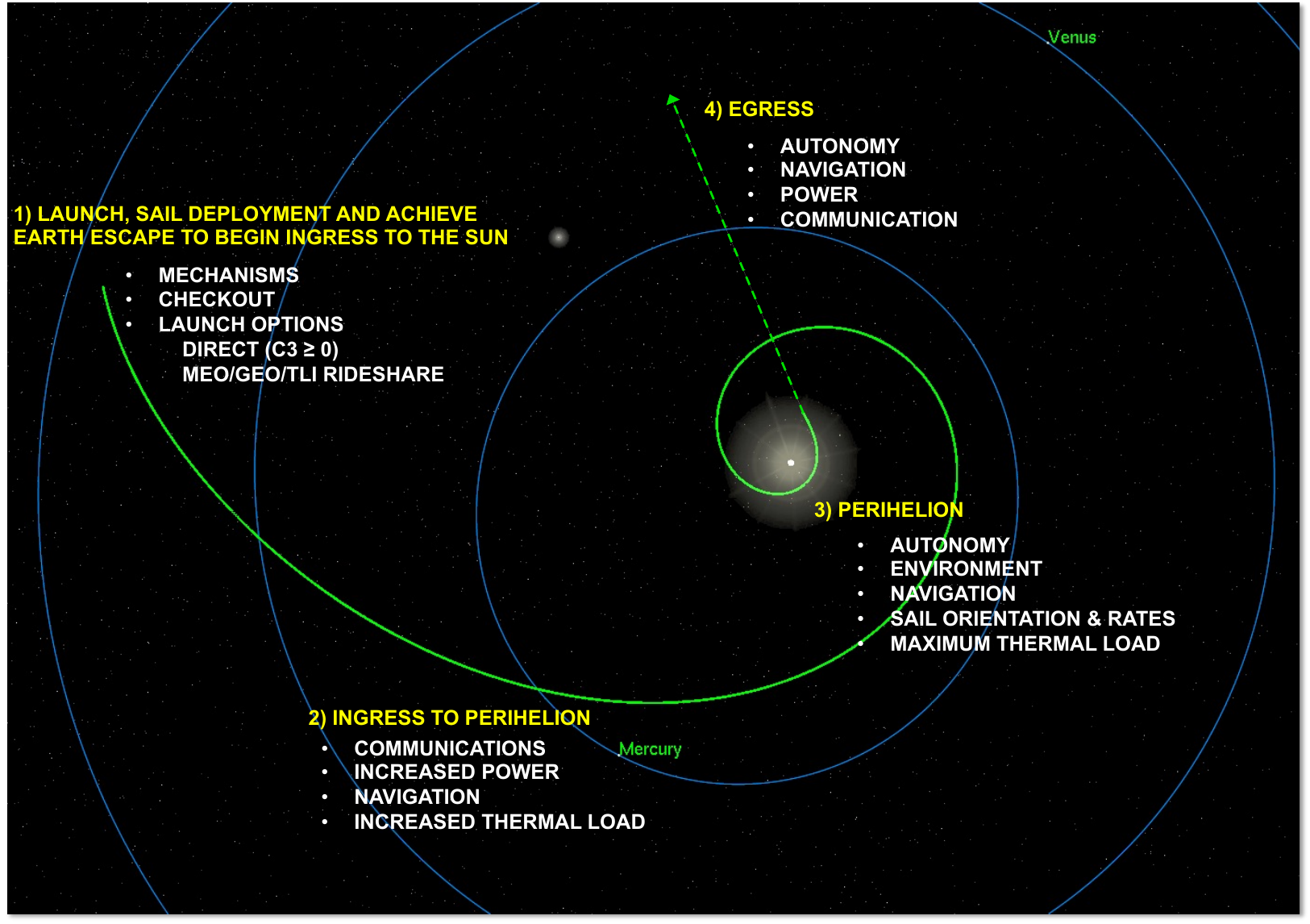}
	  \caption{ISO Intercept trajectory from Earth to solar system escape speed.}
	  \label{fig3}
\end{figure}

\begin{figure}
	\centering
		\includegraphics[scale=0.70]{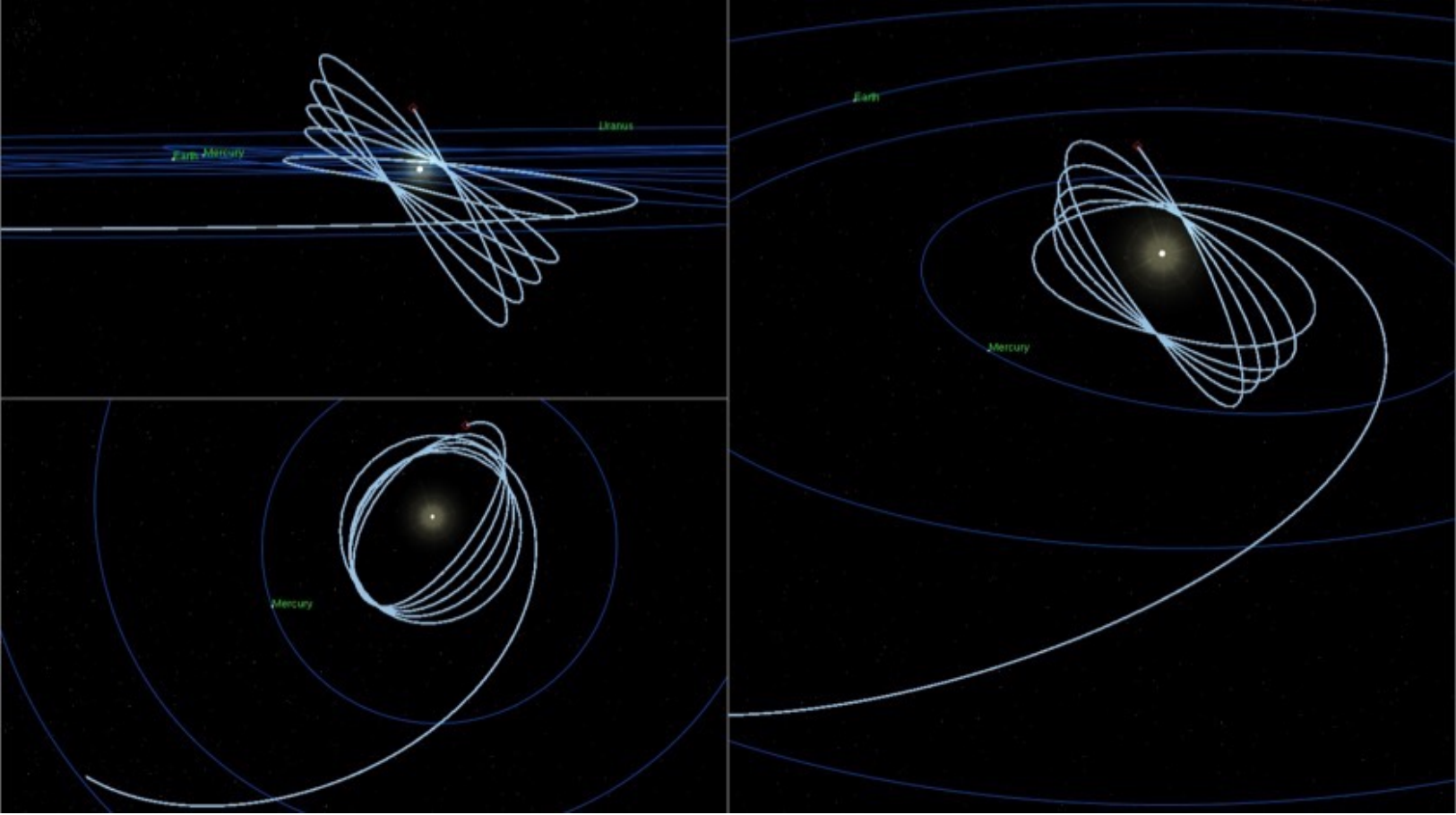}
	  \caption{Ingress to 0.2 AU perihelion and alignment of orbit to with the ISO.}
	  \label{fig4}
\end{figure}

\begin{figure}
	\centering
		\includegraphics[scale=0.70]{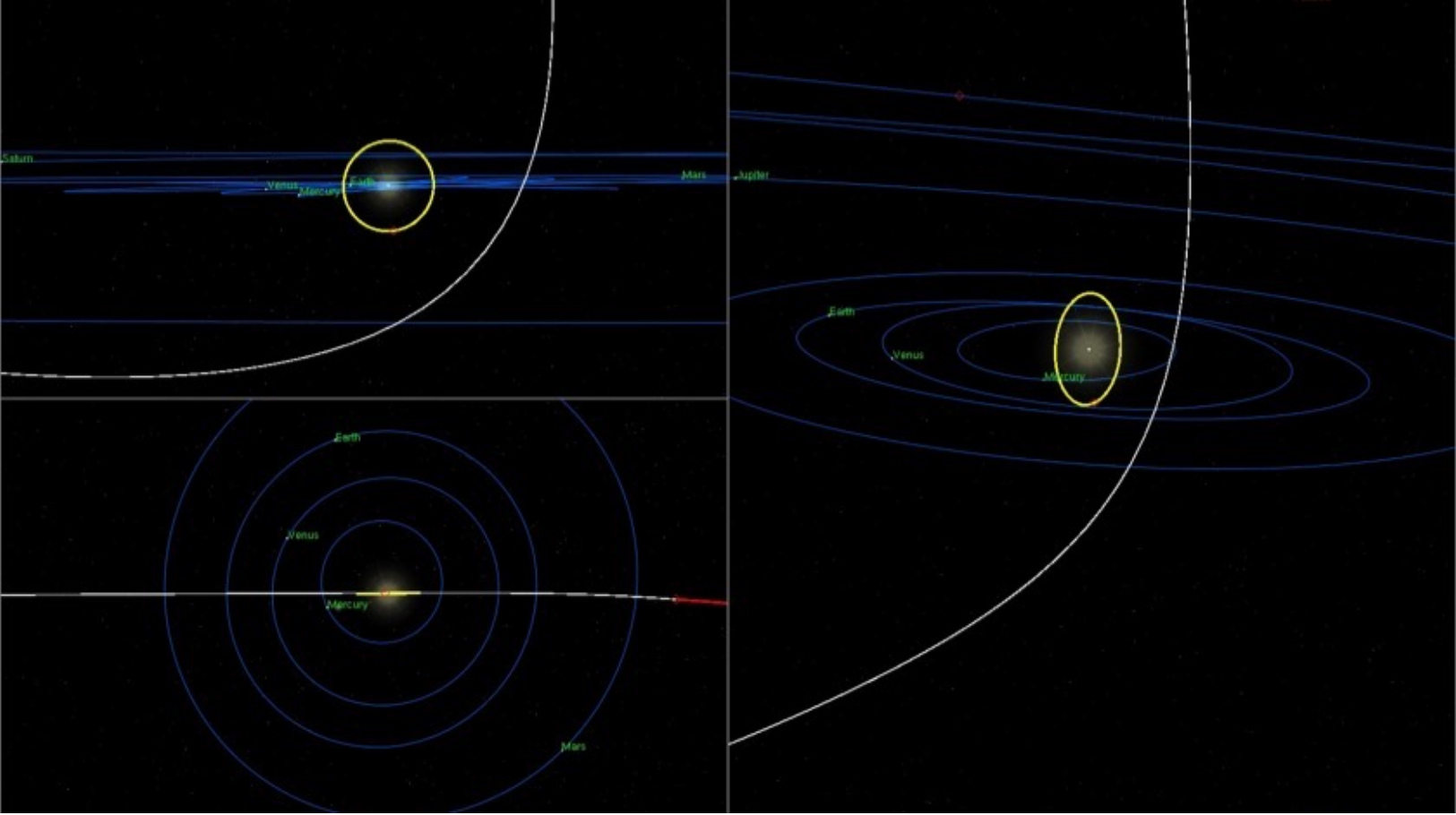}
	  \caption{Hold at 0.2 AU and phase orbit with the ISO. }
	  \label{fig5}
\end{figure}

\begin{figure}
	\centering
		\includegraphics[scale=0.31]{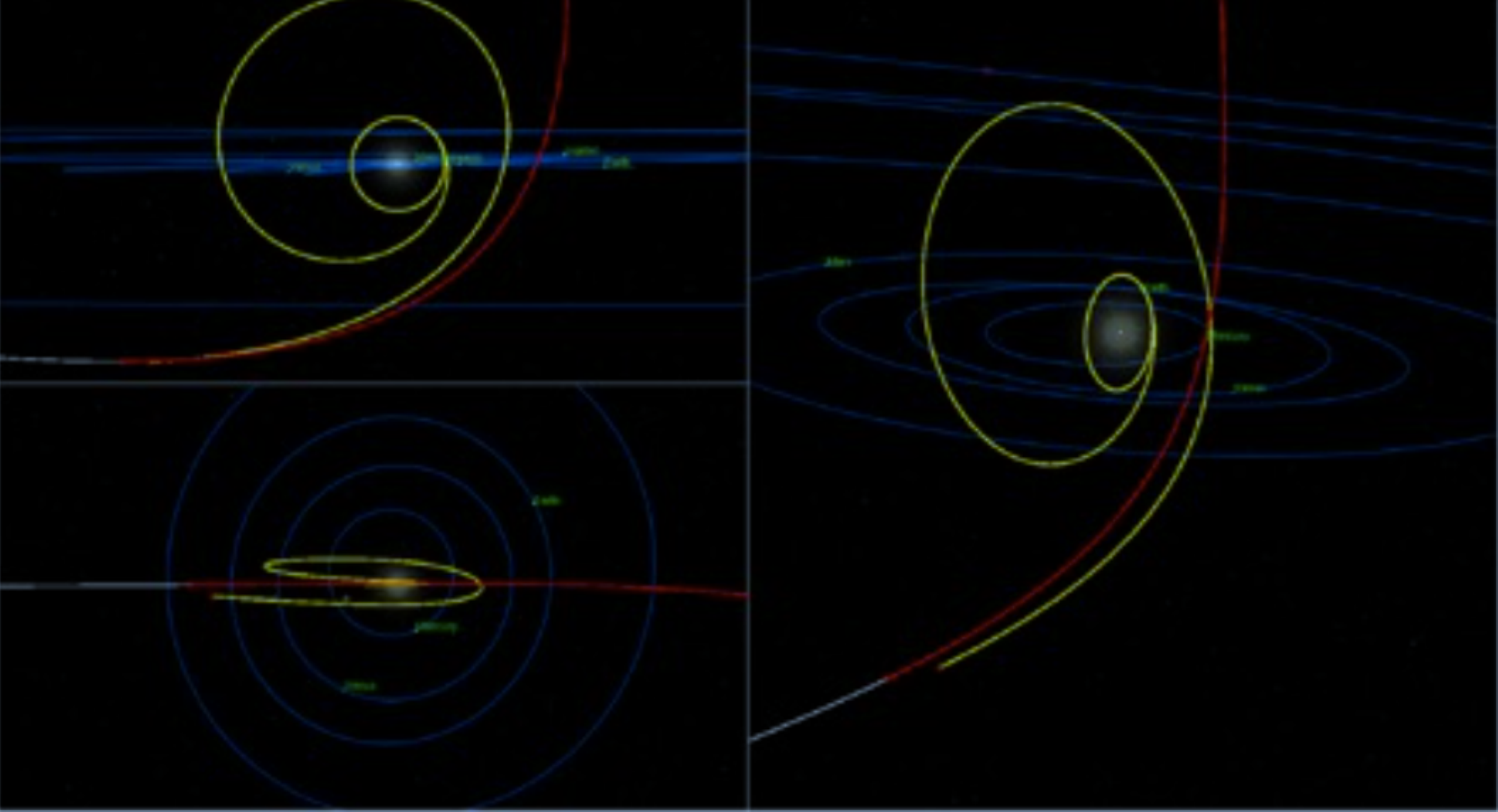}
	  \caption{Accelerate away from the Sun to catch the Interstellar Object. }
	  \label{fig6}
\end{figure}

 The low-thrust solar sail trajectory will take a few months to reach perihelion, if optimized.  But it needs not be optimized until the ISO discovery is made and can basically be ``parked'' in holding
 orbit around the Sun using the time to make solar field and particle measurements\footnote{MEMs-based devices such as radiation sensors, particle and field sensors are all ultra-lightweight instruments. Similarly, other subsystem components can be utilized for dual use experiments to include star trackers for imaging etc.}.  The spacecraft is not literally parked nor literally spiral but is maneuvered continuously with the sail to buy time until the ISO orbit is determined. Then the sail trajectory can be optimized and targeted to a perihelion chosen to maximize its resultant exit velocity and to intercept the ISO -- as shown in Figures~\ref{fig3}--\ref{fig6}, depicting the mission phases, described below:  Earth escape to perihelion, ISO orbit alignment, perihelion transfer to solar system escape velocity, approach and intercept with the ISO\footnote{Note that trajectories were generated and compared between NASA's General Mission Analysis Tool (GMAT) and Aerospace Corporation TRACE precision orbit analysis software suite.}
These mission phases can be constructed to support two distinct rendezvous profiles with vastly different mission timelines to respond to ISOs: 1) Direct Call Up: rapid response with a fast transfer from Earth (e.g. Parker Solar Probe) to perihelion to
support planetary defense and 2) Prepositioning: placing sailcraft in a holding orbit within the orbit of Mercury in anticipation of the next transient ISO. While in the holding orbit, full sun surface imaging and heliophysics can be performed as a baseline mission.

The Direct Call Up rendezvous profile has a Earth to perihelion transfer duration of weeks, while the prepositioning profile clearly illustrates is not constrained by time to reach its holding orbit nor the amount of time it maintains its close orbit of the Sun.
An example ISO intercept the prepositioning mission profile clearly illustrates the three separate phases as depicted in Figure~\ref{fig3}. The first phase is launch from the Earth to a holding orbit about the Sun. The ingress from Earth can begin with any launch but would be best served with $C_3 > 0$ launches, for instance those destined for the Moon, given the longer mission times associated with escaping Earth's gravity under sail power. Once beyond the Earth's sphere of influence, the sailcraft enters onto a heliocentric trajectory with orbital parameters controlled by sail orientation with respect to the Sun normal. During the ingress towards the Sun, the trajectory can be shaped to accommodate any launch date and remove scheduling restriction on the sailcraft deployment. Once in a holding orbit, heliophysics and space weather measurements can be made while waiting for the next transient interstellar  encounter\footnote{Alternatively, an on-demand launch from Earth can proceed after an ISO discovery, increasing the intercept timeline with an additional $\sim90$ days to account for Earth to perihelion transit time.  Another feature of the holding orbit is that inclination can be modified to meet out-of-ecliptic plane objectives.}. 

From  the holding orbit is employed, the high radiation pressure at this close distance from the Sun provides the force necessary to reorient the orbit in both inclination and longitude of the ascending node in only a few weeks after a newly discovered ISO is discovered. This reorientation is the second phase and results in aligning the holding orbit with the hyperbolic trajectory of the transient interstellar object. The final alignment is depicted in Figure~\ref{fig5}.

The third and final phase results in the sailcraft breaking from its newly aligned holding orbit and rapidly accelerating away from the Sun in a spiral trajectory that intercepts or even matches the transient interstellar object's orbit as depicted in Figure~\ref{fig6}. This allows for multiple intercepts and rendezvous scenarios.  Using smallsat spacecraft, one can think of multiple spacecraft placed in the holding orbit for multiple potential ISO discoveries. 
Using these rendezvous profiles and mission phases with multiple sailcraft enables additional advanced mission architectures to include a sample return scenario with elliptical trajectories intersecting with the ISO path, as well as planetary defense missions where a potentially hazardous object is the target of the sailcraft mission as opposed to an ISO\footnote{Additionally, one of the authors (DG) has proposed EXCALIBUR mission concept using this same scenario can be used for planetary defense -- with the potentially hazardous asteroid incoming toward Earth the target instead of the ISO.}.

Solar sail maneuverability also permits the more ambitious objective to rendezvous with the ISO -- of course with a longer trip time.  As noted, the speed after perihelion is much greater than the final hyperbolic excess speed, so interception locations within the solar system may be reached very quickly. At 0.3 AU after perihelion our heliocentric velocity is $\sim$6~AU/year.  We would catch the ISO quickly, or we could optimize the trajectory for rendezvous and potential sample return. In one scenario, an initial sailcraft intercepts or rendezvous with the ISO to perform in-situ science and to generate a debris cloud. A second, trailing sailcraft, on an elliptical trajectory, could pass through the debris cloud to collect samples and return them to Earth. Once again, fast inner solar system speeds may afford fast interception times. For instance, one sample return trajectory analyzed involves an 11-day transit from perihelion to ISO intercept near the orbit of Mercury, with an additional $\sim3.5$-year trip time to return to Earth on the elliptical return trajectory.

Finally, intercepting the ISO may require encounter operations at 5--8~AU. Using solar sail area to host solar arrays which is properly shielded near the Sun but operates without shielding in deep space may be sufficient to address all power needs. Electric propulsion micro-thrusters may be considered to add propulsion capability for maneuvering around the ISO or to provide a landing capability. This additional maneuverability may enable more detailed in-situ investigations or sample preparation activities on the surface of the ISO.

The approach to the rendezvous for articulatable vane sailcraft is  straightforward and could yield a good trajectory match. During the proximity operations where the sailcraft is approaching the ISO, data is acquired via imaging (to get size, features) and albedo (to get a sense of the mineral content). The primary goal would be to reduce the relative distance between the ISO and sailcraft. After the relative speed is sufficiently reduced, the sailcraft may land on the ISO. To enable this, electric propulsion may be used to vary the speed, while maintaining the trajectory.   Such approach could provide for sailcraft survivability and its operations at larger heliocentric distances, $\sim$ 9--10 AU. 

As a result, our concept allows for the possibility of landing on an ISO with a potentially exhaustive list of instrumentation.  The architecture employs ultrafast solar-propelled trajectories that create missions even out of the ecliptic:  An exciting objective that may revolutionize planetary science investigations.

\section{Science \& Mission Objectives}
\label{sec:sci_obj}

It is likely that many ISOs pass undetected through the solar system every year. Sending sailcraft to transient ISOs may allow us to directly access and study the building blocks of exoplanets, which could provide unprecedented constraints on planet formation models. Photographing or visiting these objects and conducting in situ exploration would allow us to learn about the conditions in other planetary systems without sending interstellar probes. The feasibility and benefits of in-situ exploration of ISO was discussed in \cite{Seligman-Laughlin:2018}.

With ISOs being distant messengers from interstellar space, we have an unprecedented opportunity to discover what these transient objects can tell us about our own solar system, its planetary formation and the interstellar medium in order to test theories regarding planetary formation.  Such discoveries may lead us to reconsider comet and planetary formation models and inform about the properties of the transient interstellar object’s parent nebula via its composition.  

A very broad range of measurements are sought for ISOs.  They include the characterization of basic physical properties (shape, density, morphology, dynamical properties), compositional properties (elemental composition, mineralogy, isotopes of at least hydrogen, oxygen, nitrogen, and carbon), geophysical/interior properties (porosity, cohesion, magnetic field), geological traits that might inform on origin and possible long-term evolution.

Simply intercepting and conducting a close flyby of an ISO is a valuable exploration objective.  Such a mission could include imaging and spectroscopic observation to estimate the size, composition, and potentially even the mass of the ISO.  This provides important astrophysical information relevant to understanding the origin and distribution of ISOs beyond the solar system \cite{Siraj-Loeb:2019}.  The mass of the spacecraft will be constrained by the desire to achieve high velocities with a high $A/m$, and so on-board instrumentation will have to carefully selected. Imaging and spectroscopy will be of primary importance.  

Given the primary requirement of achieving the necessary exit velocity via a specific area to mass ratio for the entire system, the selection and accommodation for each mission payload and subsystem is a critical design parameter. A key aspect of our ISO mission is the ability to fractionate mission payloads and functions not only across a single vehicle (e.g. utilizing the sail area for not only propulsion but also as a multifunctional structure for power generation and an antenna for communications) but also across multiple sailcraft operating in coordination to achieve the mission objectives. This additional degree of freedom with respect to mission design expands the trade space to enable larger and more massive payloads to be included by shifting other components and functions to other vehicles as a shared mission resource.

Even with the limited payload mass  it still may be possible to consider a small impactor that can be fired into the ISO to raise debris and dust for composition measurements or even for a sample return mission involving a second sailcraft that is timed to pass through the debris cloud.  As noted above, the possibilities for approaching the ISO at slow relative speed due to the nature of the low thrust trajectory and maneuverability afforded by solar sails may make it possible to rendezvous and circumnavigate the ISO for repeated observations of its topography and morphology.  Electric propulsion micro-thrusters may be considered to add propulsion capability for maneuvering around the ISO or to provide a landing capability. This additional maneuverability may enable more detailed in-situ investigations or sample preparation activities on the surface of the ISO for advanced sample return missions.  

Currently, under NIAC Phase III study \cite{Turyshev-etal:2020-Phase-III}, we are conducting a design study for series of SGL TDMs with the first of which would be the LightCraft Demonstration Mission.  A TDM is less ambitious than an ISO science mission; in fact, its payload will not be driven by science objectives and the mission only has to last about 1--2 AU beyond perihelion.  The purpose of the TDM will be to demonstrate controlling and operating the sailcraft, communicating with Earth and exercising the full capabilities of an interplanetary smallsat, and achieving a hyperbolic excess velocity faster than any spacecraft yet flown, e.g., $\sim6$~AU/year.  This work will also prove the feasibility of the ISO mission as described here.  If all goes well, we may be in a position to fly the TDM in 2024 and then conduct an ISO mission in the same decade by expanding the spacecraft platform to include the guidance, navigation, and payload instrumentation suite appropriate for an ISO mission.

\section{The Solar Sail}
\label{sec:sail}

\begin{figure}
	\centering
		\includegraphics[scale=0.80]{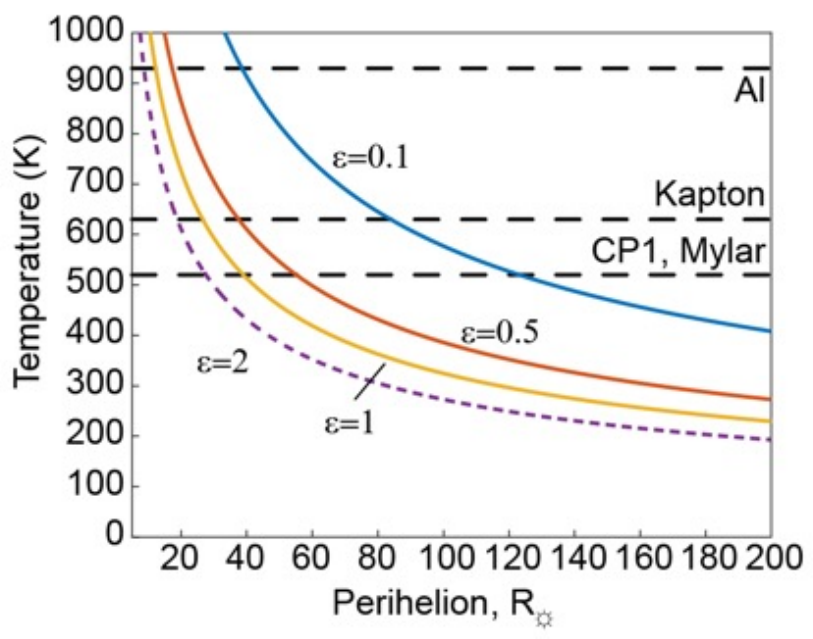}
	  \caption{Temperature as a function of solar distance for total emissivity of $\epsilon=1$.}
	  \label{fig7}
\end{figure}

The solar sail is one of the key components of the sailcraft. While a number of solar sail missions have been flown in recent years, an ISO interception mission requires passing close to the sun as described earlier. Close solar approaches may challenge sail materials and architectures. Current solar sails, such as NEA Scout and LightSail, present a thin-film, large area sail membrane that is unfurled from a CubeSat-sized platform once the sailcraft is in space. The material of the sail membrane in these instances is made of an aluminized polyimide (e.g., aluminum coated CP1\textsuperscript{TM} or Kapton\textsuperscript{TM}). Sunlight absorption by aluminum (which is $\alpha\sim10\%$) implies that the sail will be heated up. The expected sail material temperature can be predicted by calculating the thermal balance between the absorbed power and heat dissipated via thermal radiation emission into free space. Absorbed power grows inversely proportionally to the square of perihelion distance, causing the sail material temperature to increase as the sailcraft approaches the Sun. The material melting temperature then sets the limit on the closest possible approach and therefore on the maximum speed that can be attained (see Fig.~\ref{fig7}). A relatively high glass transition temperature of Kapton\textsuperscript{TM} ($\sim$600 K), as compared to other organic-based films, as well as high melting point of aluminum ($\sim$930 K) suggest that aluminized Kapton\textsuperscript{TM}  sails with a properly engineered thermal emissivity may be used for close solar approaches. Figure~\ref{fig7}  shows the sail temperature dependence of the sail material with the perihelion for different values of emissivity. From this analysis, an aluminized Kapton${}^{\rm TM}$  sail may be expected to reach as close as 0.15 AU ($\simeq 32 R_\odot$) without melting, provided that such high emissivity values are attained. One possible scenario explored to realize such a capability involves backside coating with thin-film, thermally emissive layers to further enhance Kapton's thermal emissivity. Research into advanced solar sail materials, including investigations involving novel ultra-light materials like graphene aerogel \cite{Sun-etal:2013} and other carbon-based materials, is being conducted \cite{Davoyan-2021} in collaboration with our SGLF mission study.

While an aluminized Kapton\textsuperscript{TM} sail will enable perihelion distances as close as 0.15 AU (see Figure~\ref{fig7}), other thermal challenges for the sailcraft must still be evaluated for close passes to the Sun, including the survivability of the sail interface and control mechanisms, solar cells for power, guidance and navigation systems, and communications and payload suites. A critical aspect of survivability investigations involves an exploration of heat shielding options for sailcraft elements. Here, we propose a more conservative mission scenario for initial investigations, using a perihelion of 0.2 AU.    

Manufacturing an aluminized polyimide solar sail with 2.5-micron thickness, for example, is within today's technology, as evidenced by the design of the NEA Scout solar sail.  Such a sail material has an areal density $\sim$4.1\, g/m${}^2$.  If our design goal is a $50\, {\rm kg}$ spacecraft with a total area-to-mass ratio of $A/m = 80\, {\rm m}^2/{\rm kg}$, then the sail area has to be $4000\, {\rm m}^2$.  Assuming use of a 2.5-micron aluminized polyimide sail for a $4000\, {\rm m}^2$ area, the aluminized polyimide mass would be 16.5 kg. To this will have to be added the mass of rip-stop and/or seaming, any attachments and deployment mechanisms. This should allow sufficient mass for payload and avionics to meet the $A/m$ target of $80\, {\rm m}^2/{\rm kg}$. A more advanced sail designs like the SunVane concept (see below) may enable more mass allowance for bus and payload elements, or alternatively, for even higher area-to-mass ratios that would be possible with a more traditional aluminized polyimide and stainless steel or composite boom sail design.

System design studies and trade space explorations are currently being performed to investigate area-to-mass possibilities for sailcraft as part of the TDM study \cite{Turyshev-etal:2020-Phase-III}. Preliminary estimates suggest that a reasonable design goal for the ISO mission is a 50 kg sailcraft, about 33 kg of which will be the spacecraft and its payload. If the sail could be made with thickness 2 microns, a mass savings of 3 kg would result; for a thickness of 1.25 microns, which is theoretically possible, 8 kg could be saved.  

Alternatively, the use of a new material like graphene aerogel \cite{Sun-etal:2013} may enable even more drastic mass savings. Graphene aerogel is the lightest material ever made and has a density of just $0.16\, {\rm kg/m}^3$. This may be compared with the density of polyimide ($\sim$1420$\, {\rm kg/m}^3$), aluminum ($2700\, {\rm kg/m}^3$), and carbon fiber ($\sim$2000$\, {\rm kg/m}^3$) of more traditional solar sail designs. Incidentally, aerogel graphene also features outstanding physical properties, including low thermal conductivity and temperature-invariant ($\sim$190--900$\,{}^\circ {\rm C}$) super-elasticity that allows the material to be compressed up to 90\% without permanent deformation \cite{Sun-etal:2013}. Recent research suggests the material may also be readily produced with stable properties and even 3D printed \cite{Cheng-etal:2017,Zhang-etal:2016}, all of which point to future opportunities to drastically increase area-to-mass ratios of future sailcraft.

\section{The Spacecraft}
\label{sec:scraft}

\begin{figure}
	\centering
		\includegraphics[scale=0.90]{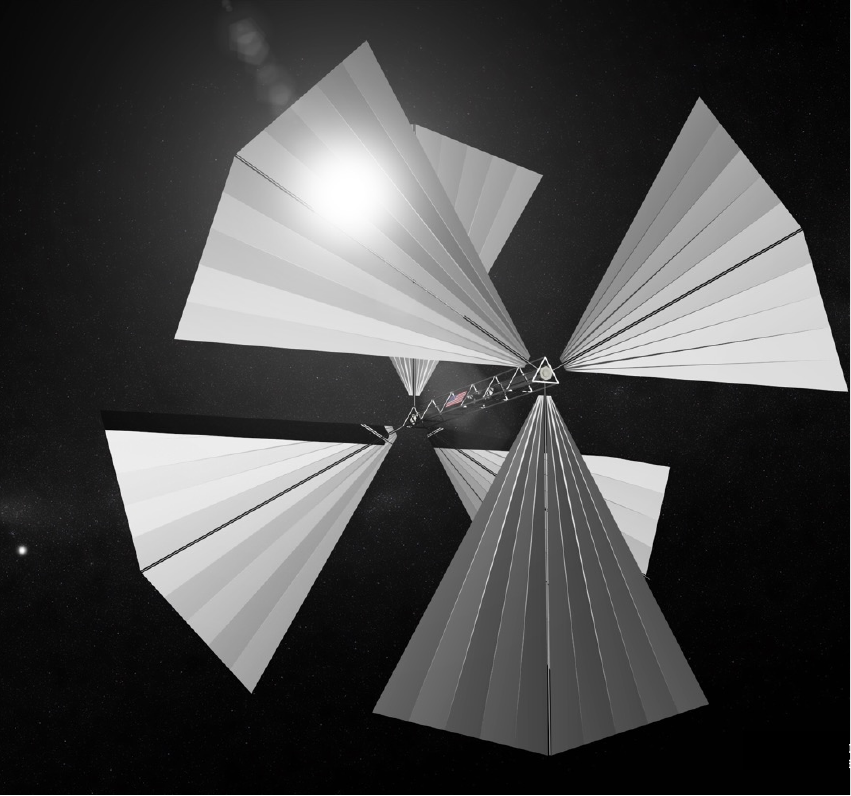}
	  \caption{The Lightcraft\textsuperscript{TM} sailcraft concept.}
	  \label{fig8}
\vskip 5pt
	\centering
		\includegraphics[scale=0.44]{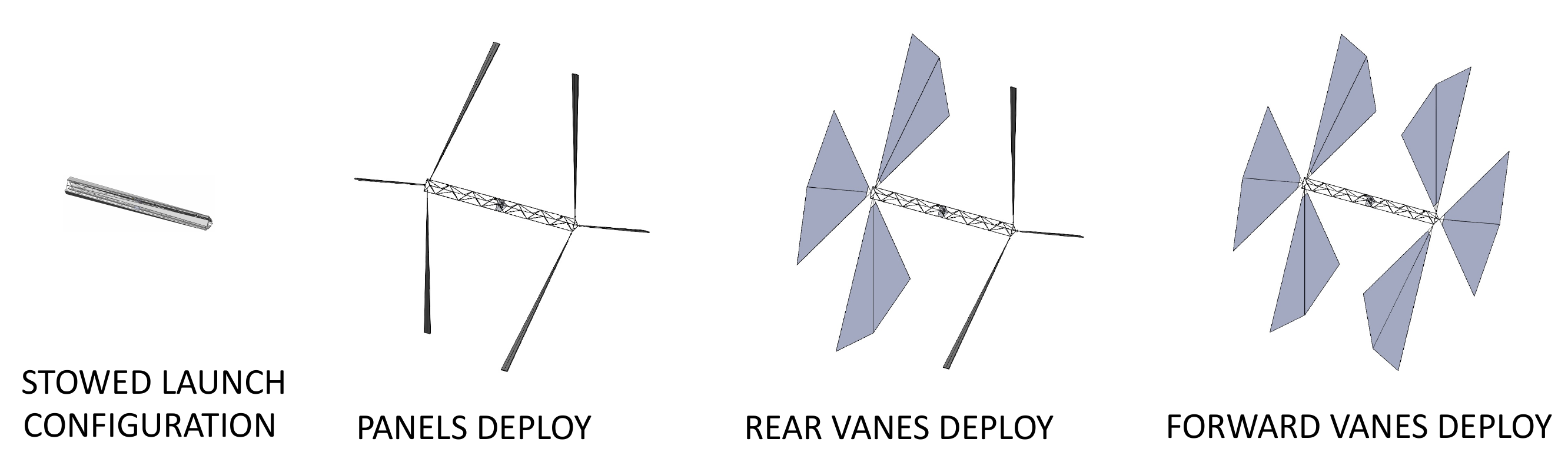}
	  \caption{Deployment and articulation of Lightcraft\textsuperscript{TM} sails. Four phases are shown: i) stowed launch configuration, ii) panel deployment,  iii) rear vanes deploy, and iv) forward vanes deploy.}
	  \label{fig9}
\end{figure}

Current solar sails (LightSail, NEA Scout \cite{Lockett-etal:2020}) are square sails.  A $4000\, {\rm m}^2$ square sail would be $\sim 63$ meters on a side.  This would be both a structural and control challenge.  Articulating such a sail may be feasible, but its dynamic stability would be hard to test.  Its control responsiveness would also be very slow, as with the IKAROS design. For this reason, a member of our team (DG), working with the designers of the L`Garde Sunjammer \cite{Sun-etal:2013,Cheng-etal:2017,Zhang-etal:2016,Aguilar-Dawdy:2000} invented a concept using the Sunjammer's control
vanes -- the small sails at corner of the square sail that would be used for attitude control. This advanced concept is now being developed by us in collaboration with L'Garde as the Lightcraft\textsuperscript{TM} to support missions from cis-lunar through to rapid exits of the solar system  (see Figure~\ref{fig8}). It positions vanes along a truss (on which the spacecraft subsystems are place) and, in principle, can be expanded to large areas with a low mass truss structure.  The revolutionary new design for solar sails (adapted from vane concept proposed for the attitude control of large square sails) provides a great deal of more controllability and eases the deployment configuration for sails.  The small size proposed makes possible a ``build and test'' approach to the sailcraft development. 

Each vane is of manageable size, can be independently controlled and articulated, and can be deployed in a straightforward procedure.  The deployment is depicted in Figure~\ref{fig9}.  For the ISO rendezvous mission, we require
an $A/m=80~{\rm m}^2/{\rm kg}$, which given mission requirements, specified payload complement and advances in miniaturization and materials allows for a wide range of sailcraft solutions spanning 10 kg spacecraft with 800 m${}^2$ of sail using cubesat components through to a 50 kg spacecraft requiring 4000 m${}^2$  of sail material. An additional trade to reduce the mass and therefore the required sail area is to distribute mission functionality across multiple sailcraft. This multidimensional trade space of sail enabled missions using this architecture is ongoing. As an example, a 15 kg integrated sailcraft with 1200 m${}^2$  total sail area could be constructed from six sails each 200 m${}^2$ in size. 

A recent concurrent engineering study done by the Concept Design Center (CDC) at the Aerospace Corporation \cite{Aguilar-Dawdy:2000} explored mission and vehicle design concepts for an ISO sample return mission based on the architecture described in this paper. A result of this study was the identification of challenges to realizing the large area-to-mass ratios that the ISO missions demand, with structures, communication, and power subsystems highlighted as specific areas that warranted further attention. Additional vehicle trade studies are also necessary to investigate multi-use or cross-use functionality for components, for instance in utilizing the large solar sail area to host solar arrays or antennas for a radio-frequency communication system.

Spacecraft power is perhaps the most critical subsystem to consider for feasibility, since power availability drives communication system performance, as well as use of on-board instrumentation and critical bus functions.  We will have plenty of solar power to at least 1.5--2 AU, and so at these ranges, thermal considerations dominate, motivating further investigations for shielding or spacecraft orientation and concepts of operations research. Intercepting the ISO will probably require encounter operations at 5--8 AU. Using solar sail area to host solar arrays that may be properly shielded near the Sun, but which could operate without shielding in deep space may be sufficient to power all necessary ISO mission needs. Even so, maintaining power for rendezvous missions in which the sailcraft remains with the ISO as it leaves the solar system will become increasingly difficult with distance from the Sun. 
An alternative that may be considered is the use of a battery system sufficiently charged for as little as 1--2 years.  This is a new way of thinking and needs quantitative estimates depending on the actual flight -- something that will be pursued for further study.   The referenced NIAC Phase II Study \cite{Turyshev-etal:2020-PhaseII} examined options for a small RTG and/or for a small battery whose charge would last through encounter -- a time of 1--2 years.  The preliminary review is encouraging, suggesting it can fit within a spacecraft design goal of 50 kg but that too reamains to be shown.

\section{Other ISO Mission Studies}
\label{sec:other-ISO-studies}

This discoveries of Oumuamua and Borisov as the first observed interstellar objects led to a number of mission proposals including Comet Interceptor \cite{Sanchez-etal:2021}, now approved by the European Space Agency (ESA) for development.  The current mission design concept calls for 2029 launch to a yet-undiscovered target. The spacecraft (yet to be designed but probably with chemical propulsion) would be launched to a Sun-Earth Lagrange point to wait for the ISO's discovery, which presumably would be within 1--2 years and thus permitting a fast flyby intercept of the ISO in the mid-2030s -- since they are designing to a flight time of approximately six years.

A solar sail mission concept somewhat similar to ours is being studied at the MIT as a student project \cite{linares2020rendezvous}.  Instead of spirling in toward the Sun and then using the close solar flyby to create the large solar force to propel the sail, they use a much higher area to mass area sail to essentially stop heliocentric velocity and produce a stasis orbit.  They propose doing this with six spacecraft placing them in different parts of the stasis orbit and then wait for the ISO discovery to then maneuver on to a rendezvous trajectory.  The orbit mechanics is innovative, but the demands on the spacecraft are far beyond any current technology.  

A mission concept called ``A Bridge to the Stars'' \cite{Moore-etal:2021} is being proposed in NASA as a New Frontiers mission.  It would also be readied for a yet to be discovered ISO and then launched within a couple of months after discovery on an Atlas V.  The spacecraft would include a deployable impactor whose impact with the ISO would be observed with camera and spectrometers on the main spacecraft as it flys very fast past the ISO.  

A general overview of the feasibility of missions to newly discovered ISOs  \cite{Hein-etal:2021} found that fast flybys were feasible now (not surprisingly)  but that rendezvous and sample return would require advanced solar electric or nuclear electric or nuclear thermal propulsion.  Unfortunately, they were unaware of the solar sailing possibilities described in this paper.  

All of these concepts involve larger spacecraft and heavier launches than we are proposing here.  Our smallsat-solar sail enables a lower cost mission that can still rendezvous with an ISO with relatively short flight time.  The scientifically interesting  low solar orbit parking orbit in our mission concept is a distinct advantage over waiting in storage on Earth or hanging out near Sun-Earth Lagrange points.   The payload capacity of a smallsat will be less than a New Frontiers or ESA-F class mission but the smallsat will still be able to carry imaging and spectrometers while the cost will be approximate one-tenth of the larger concepts.

\section{Conclusion}
\label{sec:concl}

Solar sailcraft on high-energy trajectories provide unique opportunities for intercept or even rendezvous with ISOs transiting through the solar system and may also enable even more advanced missions like ISO sample return or planetary defense. The scientific return from such investigations is invaluable, as comparative studies between an ISO sample return with solar system asteroid and comet sample returns can help us understand the conditions and processes of solar system formation and the nature of the interstellar matter.  With many new ISOs expected the significance of such an investigation is of the highest priority. 

Rendezvous or close encounter with an ISO would help us answer important questions on the formation of planetary systems including ours.
With ISOs being distant messengers from interstellar space, we have an unprecedented opportunity to discover what these transient objects can tell us about our own solar system, planetary formation, and the interstellar medium in order to test theories regarding planetary formation.  Such discoveries may lead us to reconsider comet and planetary formation models and inform about the properties of the transient interstellar object's parent nebula via its composition.   
The questions above are of high importance.  The solar sailing mission discussed here may be able to provide the key information to answer them. While future advances in materials and subsystem design open avenues for even more capable or advanced objectives, the technology needed for a basic implementation of such a mission is already available. If we so desired, we can fly an ISO mission described here within the current decade.

\section*{Acknowledgments}
\label{sec:acknowledge}

The work described here, in part, was carried out at the Jet Propulsion Laboratory, California Institute of Technology, under a contract with the National Aeronautics and Space Administration. Additional funding has been provided by the Aerospace Corporation.  The NIAC Phase III study from which this work was derived also included Breakthrough Initiatives, Xplore Inc., and Cornell Tech (a campus of Cornell University). We thank the NASA Innovative Advanced Concepts (NIAC) program for their support.  

\section*{Data Availability Statement}

No datasets were generated or analyzed during the current study.

\section*{Compliance with Ethical Standards}

The authors declare that they have no conflict of interest.

\bibliographystyle{spphys}       


\end{document}